\documentclass {article}

\usepackage{graphicx}

\author{B.M. Rodr\'{\i}guez-Lara, H.M. Moya-Cessa}
\title{Optical Bistability in a cavity with one moving mirror}

\begin{document}
\date{August, 2004}
\maketitle

\begin{abstract}
We analyze the behaviour of a coherent field driving a single mode
optical  cavity with one perfectly reflecting moving mirror and a
partially reflecting fixed mirror, and show that this system's
output exhibits optical bistability due to radiation pressure
acting over the moving mirror.
\end{abstract}

\section{Introduction}
In the  last decade cavities with moving mirrors have been
extensively studied and used to show how to synthesize
Schr\"odinger-cat states \cite{man1}\cite{bose}, prove quantum
properties of macroscopic objects \cite{bos2}\cite{gio} and more.
The experimental analysis of such a system started two decades ago
with Dorsel {\it et al} \cite{dor} who found a nonlinear behaviour
in a plane Fabry-Perot interferometer with a light mirror. Time
has passed and each day the theoretical use of this systems adds
on, but we found no quantum description for the optical
bistability phenomenon described by Dorsel {\it et al}. In this
paper we present a model intented to give a quantum description
for optical bistability observerd in cavities with one moving
mirror and relate its solution to dispersive optical bistability
in Kerr like medium. \cite{dor}

\section{Model}

We consider  a cavity with one fixed partially reflecting mirror
and one perfectly reflecting mirror, which can move in harmonic
oscillations under the influence of radiation pressure. The cavity
is coherently driven. Considering the closed system, and rotating
wave aproximation and general assumptions to evade Doppler
shifting and Casimir effects \cite{man}\cite{bose}, we are able to
write down the Hamiltonian for the closed system as:
\begin{equation}
{H}=\sum_{q=1}^{4} {H_q}
\end{equation}
Where $H_1$ is  the term corresponding to the cavity, $H_2$ the
term linked to the harmonic oscillating mirror, $H_3$ is the
coupling between the internal cavity field and the moving mirror,
and $H_4$ is the coupling between the coherent driving field and
the internal cavity field. Each one can be expressed properly as:
\cite{man}
\begin {eqnarray}
{H_1} &=& \hbar \omega_C \hat{n} \nonumber \\ {H_2} &=& \hbar
\omega_M \hat{N} \nonumber \\ {H_3} &=& - \hbar G \hat{n}
(\hat{b}^\dagger +\hat{b}) \nonumber \qquad  \\ {H_4} &=& \imath
\hbar  (\mathcal{E} e^{-\imath \omega_L t}\hat{a}^\dagger-
\mathcal{E} ^\star e^{\imath \omega_L t}\hat{a})  \nonumber
\end{eqnarray}
Where $G$ is the coupling constant for the field-mirror
interaction \cite{man1}; $\omega_C , \omega_M , \omega_L$ are the
frequencies associated to the cavity, mirror and field
respectively.

To get information about the behaviour of the output field  from
this cavity, we have to analyze the master equation for the
density operator and consider the partially reflecting mirror
coupled to a vacuum field. In a reference frame rotating at
$\omega_L$, the frequency of the coherent driving field, such open
system's master equations is \cite{car}:
\begin{equation}
\dot{\rho} = - \frac
{\imath}{h} \left[ {{H}_b,{\rho}} \right] + L_{\rho}
\end{equation}
Where the transformed Hamiltonian looks like:
\begin{equation}
{H_b}=\sum_{q=1}^{4} {H}_{bq}
\end{equation}
With components:
\begin{eqnarray}
{H}_{b1} &=& \hbar \delta \hat{n} \nonumber \\
{H}_{b2} &=& \hbar \omega_M \hat{N} \nonumber \\
{H}_{b3} &=& -\hbar G \hat{n} (\hat{b}^{\dagger} + \hat{b}) \nonumber \\
{H}_{b4} &=& \imath \hbar (\mathcal{E} \hat{a}^{\dagger}- \mathcal{E} ^\star \hat{a}) \nonumber \\
L_{\rho} &=& \gamma \left( {2\hat{a}\rho\hat{a}^{\dagger} - \rho \hat{n} - \hat{n}\rho} \right)
\end{eqnarray}
Where $\delta = \omega_C - \omega_L$ is the detunning between  the
cavity and the coherent driving field. And $L_{\rho}$ represents
the interaction with a thermal reservoir through the partially
reflecting mirror with coupling constant given by $\gamma$ or
damping rate of the cavity to the external reservoir.

With this, to ease the work we can talk about a coordinate
transformation over the total density operator:
\begin{equation}
\tilde{\rho} = D_b^{\dagger}(\kappa \hat{n}) \rho D_b(\kappa \hat{n})
\end{equation}
Where $D_b(\kappa \hat{n})= e^{\kappa \hat{n} (\hat{b}^{\dagger}
- \hat{b})} $ is the Glauber's displacement operator in mirror
coordinates, so our master equation becomes:
\begin{equation}
\dot{\tilde{\rho}} = - \frac{\imath}{h} \left[ {\tilde{H},\tilde{\rho}} \right] + L_{\tilde{\rho}}
\end{equation}
with $\tilde{H}=D^{\dagger}_b(\kappa \hat{n}){H}D_b(\kappa \hat{n})$ and expressed by:
\begin{eqnarray}
\tilde{H_1} &=& \hbar \delta \hat{n} \nonumber \\ \tilde{H_2} &=&
\hbar \omega_M \left[ \hat{N} + \kappa \hat{n}  (\hat{b}^{\dagger}
+ \hat{b}) - \kappa^2 \hat{n}^2  \right] \nonumber \\ \tilde{H_3}
&=& -\hbar G \hat{n} (\hat{b}^\dagger + \hat{b} + \kappa \hat{n})
\nonumber \\ \tilde{H_4} &=& \imath \hbar g (\mathcal{E}
D_b^{\dagger}(\kappa) \hat{a}^{\dagger}- \mathcal{E} ^\star
D_b(\kappa) \hat{a}) \nonumber \\ L_{\tilde{\rho}} &=& \gamma
\left( 2 D_b(\kappa) \hat{a} \tilde{\rho} D_b^{\dagger}( \kappa )
\hat{a}^{\dagger} - \tilde{\rho} \hat{n} - \hat{n} \tilde{\rho}
\right) \nonumber
\end{eqnarray}

In order to obtain the behaviour for the mean  field amplitude,
$\langle a \rangle$, we start from the simplyfied master equation,
transform it back to the original coordinate system and multiply
it by $\hat{a}$ by the left and trace over the result. And do the
same with $\hat{a}^{\dagger}$. It can be proved, using fundamental
trace properties \cite{bar}, that this tracing operation over the
backward transformation, $ Tr \left( {\hat{a}D_b(\kappa\hat{n})
\tilde{\rho} D^{\dagger}_b(\kappa\hat{n})} \right) $ and $ Tr
\left( {\hat{a}^{\dagger} D_b(\kappa\hat{n}) \tilde{\rho}
D^{\dagger}_b(\kappa\hat{n})} \right), $ summarize to:
\begin{eqnarray}
Tr\left( {\hat{a} D_b(\kappa\hat{n})  \tilde{\rho}
D^{\dagger}_b(\kappa\hat{n})} \right) &=& \langle \dot{a}
\rangle_{\rho}  \nonumber \\
                               &=& Tr\left( { D^{\dagger}_b(\kappa\hat{n})
                               \hat{a} D_b(\kappa\hat{n}) \tilde{\rho} }\right) \nonumber
\end{eqnarray}
With this, if we consider the initial state of the mirror as a
thermal state that will remain like that because the radiation
pressure from the field is strong enough to induce the oscillation
but not strong enough to change the state, we can express both
tracing operations as:
\begin{eqnarray}
\langle \dot{a} \rangle_{\rho}      &=& e^{\kappa^{2}  \left(
{\bar{n} - \frac{1}{2}} \right) } \langle \dot{a}
\rangle_{\tilde{\rho}}  \nonumber\\ \langle \dot{a}^{\dagger}
\rangle_{\rho} &=& e^{\kappa^{2} \left( {\bar{n} - \frac{1}{2}}
\right) } \langle \dot{a}^{\dagger} \rangle_{\tilde{\rho}}
\label{eq:mo}
\end{eqnarray}
From Eq.(\ref{eq:mo}) we notice that there's no need to  transform
back from Eq.(\ref{eq:rt}) in order to know the behaviour of the
mean field amplitude coming out from the cavity, it's enough to
analyze the cavity's creation and annihilation operators behaviour
in the transformed master equation obtained beforehand. Assuming
the density operator's form as $\tilde{\rho} =
\tilde{\rho_F}\tilde{\rho_M}$ , where $\tilde{\rho_F} = \vert
\alpha \rangle \langle \alpha \vert $ and $ \rho_M = \vert R
\rangle \langle R \vert  $, we obtain a set of Langevin equations
in the displaced coordinates:
\begin{eqnarray}
{\langle \dot{a} \rangle}_{\tilde{\rho}} &=&  g \mathcal{E}
e^{\kappa^2 (\bar{n} + \frac{1}{2}) } - \left[ {\gamma^2 + \imath
(\delta + 3 \omega_M \kappa^2  ) }\right] \langle a
\rangle_{\tilde{\rho}} - 6 \imath  \omega_M \kappa^2 \langle a
\rangle_{\tilde{\rho}} \langle a^{\dagger} \rangle_{\tilde{\rho}}
\langle a \rangle_{\tilde{\rho}}  \nonumber \\ {\langle
\dot{a}^{\dagger} \rangle}_{\tilde{\rho}} &=& g
\mathcal{E}^{\star} e^{\kappa^2 (\bar{n} + \frac{1}{2}) } - \left[
{\gamma^2 - \imath (\delta + 3 \omega_M \kappa^2  ) }\right]
\langle a^{\dagger} \rangle_{\tilde{\rho}} + 6 \imath \omega_M
\kappa^2 \langle a^{\dagger} \rangle_{\tilde{\rho}} \langle
a^{\dagger} \rangle_{\tilde{\rho}} \langle a
\rangle_{\tilde{\rho}} \nonumber \label{eq:Ls}\\
\end{eqnarray}

This Langevin set clearly shows bistable  behaviour similar to
dispersive optical bistability found in Kerr crystals
\cite{walls}. Next graphic shows the behaviour of the output field
in response to the input intensity for this system's steady state
solution

\begin{figure}[!h]
\centering \includegraphics{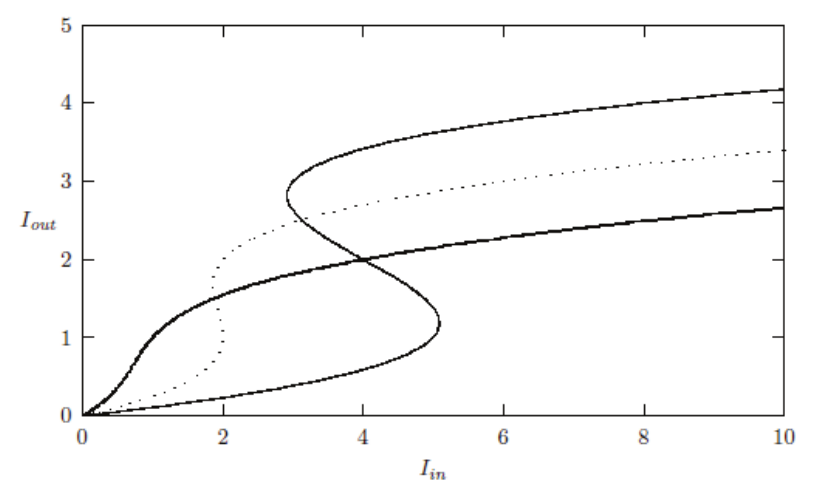} \label{fig:Lsp}
\caption{Scaled output intensity $\vert \alpha \vert ^2$  vs
scaled  input intensity $ \frac {g^2 \vert E \vert ^2 e^{\kappa^2
(\bar{n} + \frac{1}{2})}}{\gamma^2}$. With $\frac{3 \omega_M
\kappa^2}{\gamma^2} = 0.5$ and $\frac{\delta + 3 \omega_M
\chi^2}{\kappa^2} = -3.0$ (thick solid), = -2.0 (dot), = -1.0
(solid)}
\end{figure}

\section{Analysis}
Now we write down the equation that describes the semi-classical
behaviour of the output intensity of a coherently driven cavity
with one moving mirror free to oscillate under radiation pressure
from the Langevin set (\ref{eq:Ls}) and plotted in
Fig(\ref{fig:Lsp})
\begin{eqnarray}
0=g^{2}{\textbf {E}}e^{2\kappa ^{2}\left(
{\bar{n}+\frac{1}{2}}\right) } -  \gamma ^{2} {\bf a} - \left[ {
\left( {\delta + 3 \omega _{M}\kappa ^{2}}\right)  - 6 \omega _{M}
\kappa ^{2} {\bf a}}\right]^{2} {\bf a} , \label{eq:rmc}
\end{eqnarray}
and try to analyze and comment about its significance.

The quantity ${\textbf {E}}e^{2\kappa ^{2}\left( {\bar{n}+\frac{1}{2}}\right) }$
represents the scaled input field  intensity, the scaling comes first from the
coupling between the internal cavity field and the external field constant, $g$,
and then by a factor related to the initial energy of the mirror and how it is
coupled to the cavity, exponential term.

The term containing $\gamma ^{2} {\bf a}$ is related to the
transmitivity of the  partial reflecting fixed mirror in the
cavity.

Finally, $\left[ { \left( {\delta + 3 \omega _{M}\kappa
^{2}}\right)  - 6 \omega _{M}  \kappa ^{2} {\bf a}}\right]^2 {\bf
a}$ is the term that describes the action of the radiation
pressure over the moving mirror. The first term inside the bracket
would describe how the oscillation of the free mirror modifies the
detuning between the resonant frequency of the cavity and the
frequency of the input laser. The second term would describe the
direct action of the field over the moving mirror, so the full
bracket term shows how the influence of the radiation pressure
over the mirror modifies the physical length of the cavity. This
is the term that will change the resonant frequency of the cavity
and so tune the internal cavity field and the external field in
and out of resonance according to the intensity of the incident
field and so the transmitted field will increase or decrease,
respectively.

Let us take the steady state solution for the Langevin set for a
cavity with fixed mirrors containing a  Kerr-like medium of
non-linear constant $\chi$,

$$
0 = g^2 {\bf E} - \gamma^2 {\bf a} - (\delta - 2 \chi {\bf a})^2 {\bf a}
$$

Term by term analysis shows a correlation between the changes in optical length
of the cavity produced by the Kerr-like medium,  $(\delta - 2 \chi {\bf a})^2
{\bf a}$, and the change in the physical length of the cavity because of
radiation pressure, $\left[ { \left( {\delta + 3 \omega _{M}\kappa ^{2}}\right)
- 6 \omega _{M} \kappa ^{2} {\bf a}}\right]^2 {\bf a}$.

Finally, taking the classical analysis made by Meystre {\it et al} \cite{mey} for the cavity with one moving mirror
\begin{eqnarray}
I_{i}&=&I_{o}\left[ {1+R\frac{\left( {\beta _{0}+\beta _{2}I_{o}}\right) ^{2}}{T^{2}}}\right] . \nonumber
\end{eqnarray}
and comparing with our result, we can see that this pair of
equations are parallel and we can make a  similitude between
terms. This comparison is showed in Table \ref{tab:cqc}.
\begin{table}[!h]
\begin{center} 
\begin{tabular}{l l}
Classical   &   Quantum \\
\hline
$\beta_0$   &  $\delta + 3 \omega _{M}\kappa ^{2}$      \\
$\beta_2 I_o$   &  $- 6 \omega _{M} \kappa ^{2} {\bf a}$    \\
\end{tabular}
\caption[Parallel between classical and quantum analysis for a
cavity with one moviching mirror]  {Parallel between classical and
quantum analysis for a cavity with one moving mirror} \label{tab:cqc}
\end{center}
\end{table}

\section{Conclusions}
We have proposed a model to analyze the experiment performed by Dorsel {\it et
al} \cite{dor}. Our model proved practical in helping to describe, from a
quantum mechanical approach, the nonlinear response of the output field to
incident intensity, the analysis related the behaviour of this sytem to
dispersive optical bistability from fixed cavities containing Kerr-like mediums.
Finally it can be proved consistent with models for the limiting case of a
cavity with fixed mirrors.

\end{document}